\documentclass{ws-procs961x669}
\usepackage[super,compress]{cite}
\begin{document}


%
%
\def\be{\begin{equation}}
\def\ee{\end{equation}}
\def\beq{\begin{eqnarray}}
\def\eeq{\end{eqnarray}}
\def\bn{\begin{eqnarray*}}
\def\en{\end{eqnarray*}}
\def\slas{\!\!\!/}
\def\P{\Phi}
\def\p{\phi}
\def\w{\omega}
\def\W{\Omega}
\def\O{{\cal{O}}}
\def\a{\alpha}
\def\b{\beta}
\def\s{\sigma}
\def\S{\Sigma}
\def\d{\delta}
\def\D{\Delta}
\def\g{\gamma}
\def\k{\kappa}
\def\t{\theta}
\def\T{\Theta}
\def\G{\Gamma}
\def\z{\zeta}
\def\Z{\Psi}
\def\pd{\partial}
\def\e{\epsilon}
\def\n{\eta}
\def\m{\mu}
\def\r{\rho}
\def\t{\theta}
\def\R{\Rho}
\def\bra{{\langle}}
\def\ket{{\rangle}}
\def\bp{{\bf p}}
\def\bq{{\bf q}}
\def\bk{\bf k}
\def\br{{\bf r}}
\def\bx{{\bf x}}
\def\by{{\bf y}}
\def\l{\lambda}
\def\L{\Lambda}
\def\cL{{\cal{L}}}
\def\cV{{\cal{V}}}
\def\cH{{\cal{H}}}
\def\cP{{\cal{P}}}
\def\cU{{\cal{U}}}
\def\cN{{\cal{N}}}
\def\cT{{\cal{T}}}
\def\cC{{\cal{C}}}
\def\cD{{\cal{D}}}
\def\cJ{{\cal{J}}}
\def\cK{{\cal{K}}}
\def\cM{{\cal{M}}}
\def\cA{{\cal{A}}}
\def\cB{{\cal{B}}}
\def\cO{{\cal{O}}}
\def\cR{{\cal{R}}}
\def\cG{{\cal{G}}}
\def\cS{{\cal{S}}}
\def\cF{{\cal{F}}}
\def\cI{{\cal{I}}}
\def\la{\langle}
\def\ra{\rangle}

\title{Stueckelberg field and Cosmology}

\author{T R Govindarajan}

\address{
The Institute of Mathematical Sciences, Chennai 600113, Tamil Nadu, India \\
Krea University, Sricity, 517646, Andhra Pradesh, India\\
trg@imsc.res.in, govindarajan.thupil@krea.edu.in}



\begin{abstract}
Stueckelberg introduced an axion like scalar field to provide mass 
to the gauge electromgnetic field without breaking gauge invariance.
This can be considered as a precursor to the spontaneously broken 
abelian Higgs model. We will consider its role in cosmology to 
provide a novel candidate to the dark matter question. In addition 
its implications to deeper issues will be pointed out.

\end{abstract}




\section{Introduction:} Bernard Riemann introduced the axioms of geometry\cite{riemann} 
in his famous habilitation talk in 1855. This formed the basis of Riemannian 
geometry which was extended to Pseudo Riemannian geometry to provide the 
underlying structure of General relativity. But towards the end of his talk 
Riemann himself expressed his doubts about validity of his axioms in the 
infinitesimal and infinite. These difficulties, he associated with the need 
for a scale to measure infinitesimal ‘distances’ and that needs the physics of
`light rods’, essentially the wavelengths or quanta of very high frequency.
At large length scales we also need the asymptotic behaviour of the fields.
We can now associate these to the  ultraviolet and infrared divergences in the 
Quantum field theory. 

Einstein with the help of Marcell Grossmann \cite{mg14}, made his famous link between 
gravity and geometry using the axioms of Riemann. Going back to Riemann's 
observations in his habilitation talk we should relook at the implications in gravity
at these scales. We do now find the difficulties of quantising gravity and 
understanding large scale structure of galaxies can very well be linked to these 
questions.  

For the past several decades the question and origin of dark matter
\cite{dm} representing 25\% of matter in the universe have been a puzzle for 
both astrophysicists, particle physicists. The radial
speeds of luminous stars and other objects in a galaxy do not go down
sufficently fast over a large distance as expected
through Newtonian/Einsteinian  gravity. This was such an unanswered question that 
physicists started conjecturing that there could be matter within the
galaxies without any interactions to known matter but only through 
gravitational interactions. Further galaxy clusters,
bullet clusters, gravitational lensing and CMB spectrum \cite{dm} 
support this suggestion more. 

There were several candidate proposals inspired by Beyond Standard 
Model (BSM) expectations. Weakly interacting massive particles (WIMP),
SUSY particles, sterile neutrinos and many more are mentioned in the literature. 
None of them, unfortunately 
answer all the queries satisfactorily regarding this.

Situation became so desperate that even doubts were raised that gravity itself will
need changes. This led to the conjecture that law of Universal gravitation might
fail when the acceleration is below a particular threshold \cite{mond}.
This theory is known Modified theory of Newtonian dynamics (MOND). While 
by construction it explains velocity of rotational curves, it fails in
explaining bullet clusters or gravitational lensing.  In addition
it looks aribitrary without any fundamental principle behind it.

Extremely light bosons sometimes known as axion like
bosons have also been considered to account for missing mass. 
But QCD axions have not succeeded in providing such a formulation. 
But it has remained as a potential candidate within the paradigm of  
cold dark matter (CDM). Ultralight or Fuzzy dark matter is 
one the strong proposals for the dark matter.

We will explain how the Stueckelberg boson can fit in this framework in section 2
\cite{trg}
We consider the Stueckelberg boson as limiting case of abelian Higgs mechanism
in the next section (sec.3) Following this we we explain how the vortex solutions 
of such a model can play an important role in the cosmology of early universe 
leading to potential primordial blackhole and eventually super massive 
blackhole of the early universe (sec.4). Some expectations of such a proposal 
are explained in Section 5. We conclude with discussions and speculations in the end. 

\section{Stueckelberg field}
We now revisit the fundamental query raised  by Schrodinger
in 50's namely `Must the photon mass be zero?'\cite{schrodinger}.
He provided an answer and explained the details in a paper.
Since the massive photon
will have three degrees of freedom, whereas massless one will have
one less, will it affect, for example the computation of 
Stefan's constant value through blackbody radiation?  
In the computation of blackbody radiation,
we get the distribution of energy density as a function of frequencies 
and we multiply by a factor of 2 to account
for the transverse degrees of freedom. For example density of modes 
is given by:
\be
\r(\nu)~=~\frac{2h\nu^3}{c^2}\frac{1}{e^{h\nu\over kT}-1}
\label{density}
\ee
We obtain Stefan's constant by integrating the density (Eq. \ref{density})
over all the frequencies,
$$E~=~\int \r d\nu~=~ \s T^4
$$
Schrodinger himself answered the query: it will have very little effect 
if the mass of the photon is very small and if the interaction to the matter is 
through a conserved current $j^\m$. That is, if the interaction is
\begin{equation}
H_{int}~=~\int d^4x~j^\mu~A_\mu~,~~and~~\partial_\mu~j^\mu~=~0
\end{equation}
the longitudinal component will not contribute to equilibrium and escape 
through the walls (matter).
The contribution of the longitudinal photon will be negligible in the density. 
This essentially means that the longitudinal photon does not interact with matter. 
Even though photon is expected to be massless, Schrodinger adumbrated that the theory 
should make sense even if the parameter is small. He estimated the mass limit
on the photon by considering earth as a magnet and using the geomagnetic data.
He set the limit as:
\begin{equation}
m_\gamma~~\leq~~10^{-16}eV
\end{equation}
Now with more modern satellite based data, the limit has been improved as
$m_\gamma~\leq ~10^{-18}eV$.  Particle data book provides this as
the current upper bound \cite{PDG}.
\subsection{Proca and Stueckelberg theory}
Vector boson with mass, known as Proca theory has two problems
(i) It has additional degree of freedom and massless limit has a
disconituity in the number of degrees of freedom. 
(ii) Mass term explicitely breaks local gauge 
invarince. But the massive QED is renormalisable and the extra
contributions are small if the mass is small\cite{banks}. But
gauge invariance is our guiding principle and we expect it to be
preserved. Stueckelberg provided a solution that avoids simultaneously
the discontinuity in the degrees of freedom and
the absense of gauge invariance \cite{stueckelberg}.

Stueckelberg introduced an extra scalar field  $\phi$ to cancel the 
extra terms under gauge transformations and 
gave a gauge invariant massive QED action whose
massless limit in addition does not show any discontinuity.
The Lagrangian density for Stueckelberg theory is:
\be \cL~=~-\frac{1}{4} \left(F_{\mu\nu}\right)^2~+~
\frac{1}{2}m^2\left(
A_\mu~-~\frac{1}{m}\pd_\mu\phi\right)^2~+~\bar\psi\Bigl[\g^\mu(i\pd_\mu
~+~eA_\mu)~-~M\Bigr]\psi
\ee
where $\phi$ is the Stueckelberg field and $\psi$ is the matter (electron)
field. The gauge transformations are:
\be
\psi~\rightarrow e^{i\l(x)}\psi,~~A_\mu~\rightarrow A_\mu~-\pd_\mu
\l(x),~~\phi~\rightarrow \phi~+~m\l(x)
\ee
We can fix The gauge using: $\cL_{gf}~=~-~{1\over 2}(\pd_\mu A^\mu~+~m~\phi)^2$.
This can be extended to Weak and electromgnetic theory of Weinberg-
Salam model also \cite{ruegg}. 
It is well known we  can also provide mass to the photon
by the Higgs mechanism (sometimes called Anderson,Higgs mechanism)
\cite{anderson,higgs}
by coupling the abelian gauge field to a complex scalar field $\Phi$.
This will have nonzero vacuum expectation value
giving mass to the the photon. We can write in the symmetry
broken phase $\Phi~=~R~e^{i\phi}$. And the Lagrangian becomes:
\be
\cL~=~-\frac{1}{4}~F\wedge F~+~|D_\m \Phi|^2~-~V(\Phi)~+~\cdots
\ee
Phase of this field, namely $\phi$ will be the Stueckelberg field
and this mechanism in a specific limit of freezing
the fluctuations of $R$ (i.e.,make $R$ very massive)
leads to the Stueckelberg theory. Natural scale for mass of $R$ is the 
Planck mass and we will later refer to this. 
\subsection{Stueckelberg field as the dark matter candidate}
As we pointed out, Stueckelberg field does not interact
with normal matter in a
Compton length scale of $\frac{h}{m_\g c}$. But it has energy and
contributes extra terms to the energy momentum tensor. This
can take part in gravitational interaction. Can it play the role of
dark matter? But there is a problem.
Since mass of the candidate is extremely small,
they will have velocity of `light'
and decouple very early after the bigbang. During radiation  
dominated era they will not help in the structure formation.
While this was the natural expectation physics of extremely light 
massive bosons are more subtle.
At that time Bose and
Einstein save the situation. Einstein used Bose statistics 
and predicted a new phase of matter now known as Bose Einstein 
Condensate (BEC) \cite{bec}.  
Bose \cite{bose} wrote his famous paper in 1924
providing the basis for his statistics of photons. 

We propose that  we may treat the
`axion like' particles as the Stueckelberg particles
which do not interact with matter. These are also 
longitudinal photons through a gauge choice. These particles
will be such a candidate
only if they form a Bose-Einstein condensate. The relativistic Bose gas 
will form a condensate if the temperature is less than a critical temperature,
$T_c$. This temperature being $\propto \frac{1}{m_\g^3}$ is extremely 
large (close to early universe temperatures) where as the 
nonrelativisitic BEC condensate discovered in the lab 
using Rubidium gas is at nano kelvin. 
\subsection{Arguments in favor of FDM}
There are prima facie evidences for ultralight dark
matter or fuzzy dark matter proposal. Here we list some of them.
\begin{enumerate}
\item There are dwarf galaxies orbiting Milky way of
with sizes ranging from 200 light years to $10^5~$ light years.
There are about 1000 to $10^6$ stars in them. In some, stars are at the
initial stages of formation. Dark matter content is more in these
dwarf galaxies. N body Simulation with usual dark matter candidates
indicate the density grows exponentially at the core. This is known as 
core cusp problem \cite{corecusp}. But
for fuzzy dark models there is no such problem.
This is because FDM behaves like a fluid or wave. About 100 to 200
such galaxies are expected through simulations. 
But only about 40 such galaxies have been
seen so far. There are many more expected but are probably too faint
and/or DM content is too high and hence less luminous.
\item There are  dwarfs
with sizes less than 1000 light years. The compton wavelength
of the Stueckelberg particles will range from few light days to
100-150 light years depending on the mass from $10^{-20}~-~10^{-24}
eV$. The size of the BEC condensate can be expected to be
that of compton wavelength i.e.,
$\cO\left(\frac{h}{m_\g c}\right)$. The size of smallest galaxy will
set a limit on the
compton wavlength and hence of the mass of the dark matter candidate.
The smallest dwarf galaxy known so far is Segue-2 
with a half radius of 115 light years, at a disatnce of $10^{5}$ light
years away \cite{segue2}. It has a mass of
$\cO(10^6)M_{\odot}$ with around 1000 stars. It has size of Ursa Minor! 
We can expect 100 light years as the limit set by dwarf galaxy sizes
which will correspond to $m_\g \geq 10^{-24}eV$.
\item Ryutov {\em et al}  showed the Maxwell Proca electrodynmaics
changes stress tensor due to longitudinal photons. This stress under certain
conditions develope a `negative pressure'\cite{ryutov}.
The effect is associated with random
magnetic fields with correlation lengths exceeding
the photon Compton wavelength. 
This is precisely the scale at which the longitudinal
photons contribute. But unfortunately this  force
is not sufficient to explain the contribution to a star like sun
in the rotational curves. This might be because we evaluate contributions in
the Proca theory. We need to  study this theory 
in the context of spontaneous symmetry breaking through
Higgs mechanism. See in this connection Dvali etal \cite{dvali}.
In this connection, an interesting observation is made about the photon mass 
by Alessandro Spallicci etal., \cite{spallicci}. 
In the standard model extension with Lorentz symmetry 
violation gives dressed photon without breaking gauge invariance. 
This photon mass induces an effective dark energy which acts optically, 
but not dynamically. This can help in resolving Hubble tension 
through SNeIa distance and luminosity. We will
comment on this in future publication. 
\item Recently results from Pulsar Timing Array (PTA) is providing certain restrictions 
on the scenario of the FDM. Pulsar timing array is a collection of radio telescopes looking
for Nano hertz gravitational waves due to merger of galaxies
in the very early universe. By analysing the
time period of the Pulsars within our Milkyway galaxy.
they have detected recently with confidence level {98\%} a
hum of the early universe.
This is similar to CMB for Electromagntic spectrum.
It also has an interesting connection to FDM.
The gravitional waves of the galaxy mergers of the
early universe are affected by the dark matter between pulsars
in our galaxy. The average distances
of Pulsars should be comparable to the Debroglie wavelength of FDM
candidates. We expect the mass of FDM
candidate should be $\geq 10^{-24}eV$ \cite{rubakov}.
Parkes Pulsar Timing array which has been monitoring
20 millisecond pulsars data for the period 2004-16 has been studied.
A limit FDM mass of $\geq 10^{-24}$ is claimed.
\end{enumerate}
\section{Abelian Higgs Mechanism}
Stueckelberg model as pointed out earlier,  does not break gauge invariance
and has no discontinuity  in the degrees of freedom when the mass parameter
is made zero. It can also be obtained as a suitable limit of
the abelian Higgs mechanism wherein the $U(1)$ gauge field is coupled to
a complex scalar field with spontaneous symmetry breaking.
We will elaborate this now and see how it can provide a mechanism for
seeds of primordial blackholes which can grow as supermassive
blackholes.

Consider a complex scalar $\Phi$ coupled to $U(1)$ gauge theory. The action
can be written as
\bn
S~&=&\int \cL~d^4x~=~\int ~d^4x~\left(-
\frac{1}{4}F_{\m\nu}F^{\m\nu}~+~|D_\m\Phi|^2~-~V(\Phi)\right)\\
V(\Phi)~&=&~-\m^2~|\Phi|^2~+~\l~|\Phi|^4
\en
Here $\l$ is quartic coupling constant.
The potential is a double well potential and gauge symmetry will be broken
with $|\Phi|^2~=~v^2~=~\frac{\m^2}{2\l}$,  
and $v$ is the expectation value of scalar field at the ground state.

Expanding the field
as $\Phi~=~\sqrt{\frac{\m^2}{2\l}}~e^{\frac{i\p}{v}}$ we will get,
\be
\cL~=~-\frac{1}{4}\left(F_{\m\nu}\right)^2~+~m^2\left(A_\m~-~\frac{1}{m}\pd_\m \p
\right)^2~+~\cV(v,\p)
\ee
where $m~=ev$ is the mass of the photon. We can make the Higgs mass very high
(as it is not naturally protected). it can be as high as $\cO(10^{16}) GeV$
and charge extremely small so that photon mass is finite and small value $\leq
\cO(10^{18})eV$. It is easy to see the Stueckelberg model emerging in the 
limit of infinite mass of the Higgs. This is the scenario at the very 
early Universe. The dark matter condensate
should be formed in the radiation era with the Stueckelberg scalars.
The condensate size can be expected to be $\cO({h}/{m_\g c})$.
To be precise we should use Debroglie wavlength at the high temperature.
But that will probably alter the estimate only by an order of magnitude.
This length scale for $m_\g~=~\cO(10^{-20})eV$ is $\approx~ .1$ light year.
For $m_\g~=~\cO(10^{-23})eV$ it is 100 lightyears. Since we expect
from the minimum of the dwarf galaxy the mass to be in this range we can
ask how long will it take for the Universe size to grow to 100 light years
through Friedman metric. This is estimated to be about 1000 to 10000 years.
This is the time scale for ultralight dark matter condensate to be formed.
Given this estimate from the Stueckelberg field we can look for the role
of the Higgs part of the complex scalar field. Its mass is very large and
acts as the seed of the primordial blackhole.
\section{Vortex solutions of Abelian Higgs model}
We now should consider fluctuations of the heavy Higgs also. With 
this  the Hamiltonian can be written as:
\be 
H~=~\int d^3x \frac{1}{2}\left(E^2+B^2 + m^2~\cA^2\right) ~+~\cV(v, \phi)
\ee
Now $\cA$ is in addition to transverse polarisations of the electromagnetic field 
contains the longitudinal polarisation too. Now generally we ignore the fluctuations 
of the Higgs field since the mass of Higgs is very large. But over length scales of 
compton wavelength of photons one can lower the energy by creating localised zeroes 
of the Higgs field. This will be like Abrikosov vortices in type II superconductor
\cite{vilenkin}.
The phase of the Higgs field can compensate the energy due to the transverse component.
Actually this is due to the topological winding around the zeroes of the Higgs field 
due to the longitudinal component.  There is critical field $B_c$ above which photon 
behaves like a massless field. But $B < B_c$ it has mass and behaves like massive 
proca theory. 

But $B_c$ is  very large since the expected mass of the photon is very small. 
That means a large number of vortices should be present to compensate and lower 
the energy configuration. 

For $m_\g$ is $\cO(10^{-20})$ eV and charge $e~=~10^{-19}$ we can have high mass Higgs
which is not unnatural. This can lead to simultaneous presence of ultralight dark matter as 
condensate and large mass Higgs as seeds of primordial black holes.
The condensates can be the the size of compton wavelength of stueckelberg quanta.
The condensate can also be accreted by the primordial black hole to develope into  
supermassive blackhole earlier than expected. This scenario provides 
simultaneouly dark matter and supermassive blackhole aiding the structure formation
like galaxy. This outline of the program requires further details to make suitable 
predictions. 
\section{Conclusions}
Photon mass being zero has been questioned earlier \cite{debroglie}.
The question of combining gauge invariance, Lorentz invariance for such a 
theory has led to ideas of spontaneous symmetry breaking and provided 
interesting connections to the cosmology. 
We would like to caution about the status of photon mass limits \cite{spallicci1}
which will be crucial for further developements.
We have proposed a novel candidate for the ultralight dark matter. It is the Stueckelberg
quanta which behaves like an axion like scalar particle at length scales less than 
its compton wavelength. This provides a potential limit of the mass of photon between
$\cO(10^{-18}) - \cO(10^{-24})$ eV corresponding to compton wavelength of few light seconds 
to 200 light years. This fits the size of the smallest dwarf galaxy seen so far 
and has the potential of providing seeds for primordial blackhole. Formation 
of BEC condensate plays crucial role in this. The seeds 
itself can accret dark matter condensate  and   grow as supermassive blackhole in 300 - 400
million years after the big bang. Evidences for galaxies at such early times is being obtained
by JWST and needs a good theory to explain. These require further detailed study of 
the first order phase transition of Abelian Higgs model at the cosmological scales 
in the early universe\cite{gravityprize,mg17,zeldovich}. This will be followed up later.

We go back to profound remarks due to Bernhard
Riemann on the ultimate structure of space time in his 
Habilitation talk \cite{riemann}.  

\begin{center}{\bf
AXIOMS UNDERLYING THE BASIS OF GEOMETRY \\
MAY NEED CHANGES\\
AT INFINETESIMAL AND INFINITE LEVEL.\\
Bernhard Riemann}
\end{center}

These changes will be inherently tied up with quantum nature of gravity. 
\section*{Acknowledgements}
I acknowledge Krea University for providing partial support for taking part in the 
conference. I thank Dr Sivakumar, Dean of Research, Krea University for facilitating.
I thank the referee for meticulously correcting many of my references and providing 
new informations. 

\end{document}